\documentstyle[twocolumn,aps,prd,floats,epsf]{revtex}

\tightenlines

\newcommand{\be}{\begin{equation}}
\newcommand{\ee}{\end{equation}}
\begin{document}

\renewcommand{\topfraction}{0.99}\renewcommand{\bottomfraction}{0.99}
\twocolumn[\hsize\textwidth\columnwidth\hsize\csname@twocolumnfalse\endcsname

\title{Attractive Forces Between Global Monopoles and Domain Walls} 
\author{Stephon Alexander,\footnotemark[1]
 Robert Brandenberger, and 
Richard Easther\footnotemark[2]}  
\address{ Physics Department, Brown University,
Providence, RI 02912, USA}

\maketitle

\begin{abstract}
We study the interaction between stable monopoles and domain walls in
  a $SO(3) \times Z(2)$ scalar field theory. Numerical simulations
  reveal that there is an attractive force between the monopole and
  the wall, but that after the monopole and the wall collide, the
  monopole does not unwind.  We present an analytic explanation for
  the origin of the attractive force, and conclude that this is a
  generic feature of monopole-wall interactions which does not depend
  on the detailed structure of the model.  The existence of the
  attractive force supports the hypothesis of Dvali {\em et al.\/}
  (hep-ph/9710301), who proposed that monopoles can be ``swept up'' by
  domain walls, thereby alleviating or solving the monopole problem
  associated with phase transitions occurring after or in the absence
  of inflation.
\end{abstract}

\raggedleft{BROWN-HET-1207 }

\mbox{}]

\footnotetext{\footnotemark[1] 
  After September 1, 2000: Imperial College, London.}
\footnotetext{\footnotemark[2] 
  After October 1, 2000: Columbia University, New York.}

\section{Introduction}

In a previous paper in collaboration with Andrew Sornborger, we
investigated the interaction of global topologically stable monopoles
with embedded domain walls in a $SO(3)\times Z_2$ linear sigma model
\cite{ABES}.  We were motivated by Dvali {\em et al.\/}'s proposal
\cite{DLV} that monopoles could be swept up by domain walls which, if
true, provides a potential solution for the monopole problem.  In our
previous analysis the $Z_{2}$ symmetry factor corresponding to the
domain wall was a ``semi-discrete'' symmetry, i.e.  part of the
connected symmetry group.  Hence, the walls corresponding to this
$Z_2$ were not stable; they were {\it embedded} walls
\cite{embedded,AV99}. However, we confirmed that both monopoles and
anti-monopoles are attracted to the domain wall and that after
colliding with the wall they unwind before the wall itself
decays. Furthermore, we confirmed Dvali {\em et al.\/}'s conjecture
that the monopole charge dissipates on the surface of the wall by
studying the evolution of both the winding number $Q_{w}$, and the
winding number density $\rho_{w}$.

The question immediately prompted by our previous results is whether
the same situation will be found when global monopoles interact
with topologically stable domain walls.  Recent work \cite{PV}
provides numerical evidence that the attraction-unwinding mechanism
extends to models with both topologically stable local monopoles and
domain walls. In this paper, we ask whether or not the
attraction-unwinding mechanism carries over to models with stable
global monopoles and $Z_{2}$ walls.

In our previous work \cite{ABES}, we argued that the attractive force
between the monopole and the wall would also be found in systems with
general global and gauge monopoles and domain walls because the
underlying physics is generic, and rests primarily on the minimization
of the gradient energies of the field configurations.  In this paper
we present evidence supporting this conjecture by studying a $G =
O(3)\times Z_{2}$ linear sigma model spontaneously broken to $H =
O(2)$. The vacuum manifold is $\cal{M} = G/H =\rm S^{2} \times Z_{2}$
and thus admits both topologically stable domain walls and monopoles.
Our numerical simulations confirm the existence of the attractive
force between monopole and domain wall.  However, in contrast to the
interaction between global monopoles and embedded global walls, we
find that the monopole fails to unwind on the surface of the domain
wall. Finally, we provide an analytic explanation of the results seen
in the numerical simulations.

\section{The Model}

Topologically stable monopoles are ensured \cite{Kibble76} when the
second homotopy group of the vacuum manifold is nontrivial, i.e.
$\Pi_{2}(\cal{M} \rm) \neq 0$. Similarly, stable domain walls require
the existence of a nontrivial $\Pi_0(\cal{M})$. Therefore, to get both
stable domain walls and monopoles one needs $Z_{2}$ and $S^{2}$
factors to be present in the vacuum manifold.  In this investigation
we consider a field theory with a novel potential which admits just
such a vacuum manifold. The symmetry breaking pattern is 
\be 
G = O(3) \times
Z_{2} \rightarrow H = O(2) \, .  
\ee 
Therefore, the vacuum manifold is
\be {\cal M} = \frac{G}{H} = \frac{O(3) \times Z_{2}}{SO(2)} \simeq
S^{2} \times Z_{2} \, .  
\ee

To realize a symmetry breaking pattern which yields both stable
monopoles and domain walls, we use \cite{Unruh} the following
Lagrangian: 
\begin{eqnarray}
{\cal L} &=& \frac{1}{2} \partial_{\mu}\Phi^{a}\partial^{\mu}\Phi^{a} 
- \frac{\lambda_{1}}{4}\phi_{4}^{2}(\phi_{1}^{2} + \phi_{2}^{2}+
\phi_{3}^{2} - \eta_{1}^{2})^{2} \nonumber \\
&& - \frac{\lambda_{2}}{4}(\phi_{4}^{2}-\eta_{2}^2)^{2} \label{pot}
\end{eqnarray}
where $\Phi^{a}= (\phi_{1},\phi_{2},\phi_{3},\phi_{4})$ is a four
component scalar field.
 
This model has an exact $Z_{2}$ symmetry corresponding to the discrete
transformation $\phi_4 \rightarrow - \phi_4$. Furthermore, the
Lagrangian is invariant under an $O(3)$ rotation of the fields
$\phi_{1}, \phi_{2}, \phi_{3}$. Hence the Lagrangian has an $O(3)
\times Z_{2}$ symmetry. The set of vacuum states is
\begin{eqnarray}
\phi_1^2 + \phi_2^2 + \phi_3^2 &=& \eta_1^2 \nonumber \\
\phi_4^2 &=& \eta_2^2
\end{eqnarray}
and the vacuum manifold hence is $S^2 \times Z_{2}$.

Since 
\be
\Pi_{2}({\cal M})= \Pi_{2}(G/H) = Z \times Z_{2} \neq 1 \, ,
\ee
the existence of stable monopoles is ensured. A monopole configuration is
\begin{eqnarray}
\phi \, &=& \, (\phi_1, \phi_2, \phi_3) = f(r,t) (\frac{x}{r},
\frac{y}{r}, \frac{z}{r})\eta_1 \nonumber \\
\label{mon} \phi_4 \, &=& \, \pm \eta_2 
\end{eqnarray}  
where $r^2 = x^2 + y^2 + z^2$ and where the profile function $f(r,t)$
obeys $f(r,t) = f(r) \rightarrow 0$ as $ r \rightarrow 0$ and $f(r,t)
= f(r) \rightarrow 1 $ as $ r \rightarrow \infty$.  A reasonable
ansatz for the profile function at the initial time $t_0$ is 
\be
f(r,t_{o}) \, = \, (1-e^{- \frac{r}{r_c}}) \, , \label{eq5}
\ee
where $r_c$ is the core radius of the monopole which is determined by
balancing gradient and potential energies, which yields $r_c \sim
\lambda^{-1/2} \eta^{-1}_{1}$.

Likewise, since 
\be
\Pi_{0}(\cal{M} \rm) \neq 0 \, ,
\ee
our Lagrangian also supports the existence of stable domain walls.  A
domain wall is constructed by picking one spatial direction (e.g. the
$z$ direction), and setting up a kink configuration in the $\phi_4$
field: 
\be \label{wall} \phi_4(z) \, = \, \eta_2 \tanh\left(
\frac{z}{w}\right) \, ,
\ee
and letting the other three fields be at a fixed point in the $S^{2}$
subspace of $\cal{M}$. Here, $w$ is the thickness of the wall,
$\sqrt{2 /\lambda_{2}}/ \eta_{2}$.

This model has several advantages compared to the model with three
scalar fields studied in \cite{ABES}. In particular, it admits stable
domain walls rather than unstable embedded walls. Moreover, it is much
easier to set up a configuration corresponding to a monopole displaced
a certain distance $z_0$ from a domain wall as one does not have to
rely on a product ansatz to simulate the monopole / domain wall
configuration. The $\phi_{4}$ field independently generates a pure
domain wall while the other three fields will support the monopole
solution without affecting its winding number. Specifically, we can
start with the $\phi_4$ field in the configuration (\ref{wall}) and
set up the other three fields in the configuration (\ref{mon})
centered not at $(x,y,z) = (0,0,0)$, but at $(x,y,z) = (0,0,z_0)$.
The mixing of the winding number due to the product ansatz was a
concern in our previous work \cite{ABES}, and now we can address it
explicitly.

We also need to study the monopole winding number, $Q_{w}$, in order
to determine whether the monopole unwinds.  We can compute both the
winding number and the winding number density.  The topological charge
associated with $\Pi_2({\cal M})$ (which in our case is ${\cal Z}$) is
the winding number. This number quantifies how many times the field
configuration $\phi(x)$ wraps around the vacuum manifold as $x$ ranges
over a sphere $S^{2}$ in coordinate space. Hence the winding is
defined as the homotopy class of the map $\hat{\phi}=
\frac{\phi^a}{|\phi|}$ from coordinate space to the vacuum manifold,
known as isospace: $\hat{\phi} :S_{space}^{2} \rightarrow
S_{iso}^{2}$.

For our O(3) theory, the winding number in a volume enclosed by the surface 
$S_k$ is:
\be
N= \frac{1}{8\pi} \oint
dS_{k}\epsilon^{ijk}\epsilon_{abc}\hat{\phi^a}\partial_i\hat{\phi^b}
\partial_j\hat{\phi^c} \, , 
\label{wind} 
\ee  
where the indices $a,b,c$ run from 1 to 3.

Since we consider the time evolution of the winding over the entire
coordinate space, useful information is obtained from the winding
number density.  Using Stokes' theorem in Eq. \ref{wind} by performing
a total derivative on the surface flux in the integrand, the winding
number becomes:
\be N= \frac{1}{8\pi} \int
d^{3}x\epsilon_{abc}\epsilon^{ijk}\partial_i\hat{\phi^a}\partial_j
\hat{\phi^b}\partial_k\hat{\phi^c}.
\label{wind2}
\ee
By visualizing the evolution of the integrand in Eq. \ref{wind2} we
obtain information about the topological charge density over the whole
space.  In principle, we could track the charge by studying the
surface flux alone, but to do this one has to know where the winding
is in order to choose a correct surface of integration, or compute the
surface integral over many different trial boundaries.  Consequently,
we have found the winding number density to be a more useful variable.

The one drawback of using the winding density to track the evolution
of the monopole winding is that the density is initially a delta
function, since $\epsilon_{abc}\epsilon^{ijk}\partial_i\hat{\phi^a}
\partial_j\hat{\phi^b}\partial_k\hat{\phi^c} \sim \delta(r)$ and the
delta function is ill defined after we discretize the equations of
motion.  Consequently, while the observed peak in the winding density
will give an accurate qualitative picture of the monopole's location,
the volume integral Eq. \ref{wind2} is not conserved during the
numerical evolution, even when we are looking at an isolated monopole.

\section{Results and Discussion}

Partly as a test of the code, we first investigated the monopole and
domain wall in isolation.  In both cases we found that the solitons
are stable.  Note that if the initial thickness of the solitons is not
chosen ideally, there will be excess energy which can and will radiate
away. Due to the reflective boundary conditions implemented, this
energy reflects back towards the core. Nonetheless, in our studies of
monopole - wall interactions this was not an obstruction, since the
interaction occurred before the boundary effects became significant.

Our primary goal was to investigate the attractive force between the
monopole and wall.  We did this by choosing initial conditions
corresponding to a monopole that was initially at rest adjacent to a
domain wall, and then numerically evolving the three dimensional
equations of motion.  The evolution of the energy and winding
densities for a typical set of parameters is displayed in Figs 1 and
2. The parameter choices for the plotted solution are $\lambda_1 =
\eta_1 = \eta_2 = 1$ and $\lambda_2 = 2$, and the equations of motion
were solved on a $192^3$ point grid. 

The attractive force between the monopole and wall is obvious from the
simulation, and does not depend on any special parameter choices. In
Fig. 1, we see that the monopole starts moving towards the domain
wall, but does not pass through it. At later times (not shown in the
plots) the distance between the monopole center and the wall does not
decrease uniformly, but oscillates.  This is apparently due to the
natural oscillations of the monopole that are observed in isolated
monopoles if the initial profile is not a time independent solution,
and may possibly be accentuated by the edge effects due to the finite
size of the simulation volume. However, there is no ambiguity about
the existence of the attractive force that leads to the initial
displacement of the monopole.

\begin{figure}[htbp] \begin{center}
\begin{tabular}{c}
\epsfysize=4.5cm 
\epsfbox{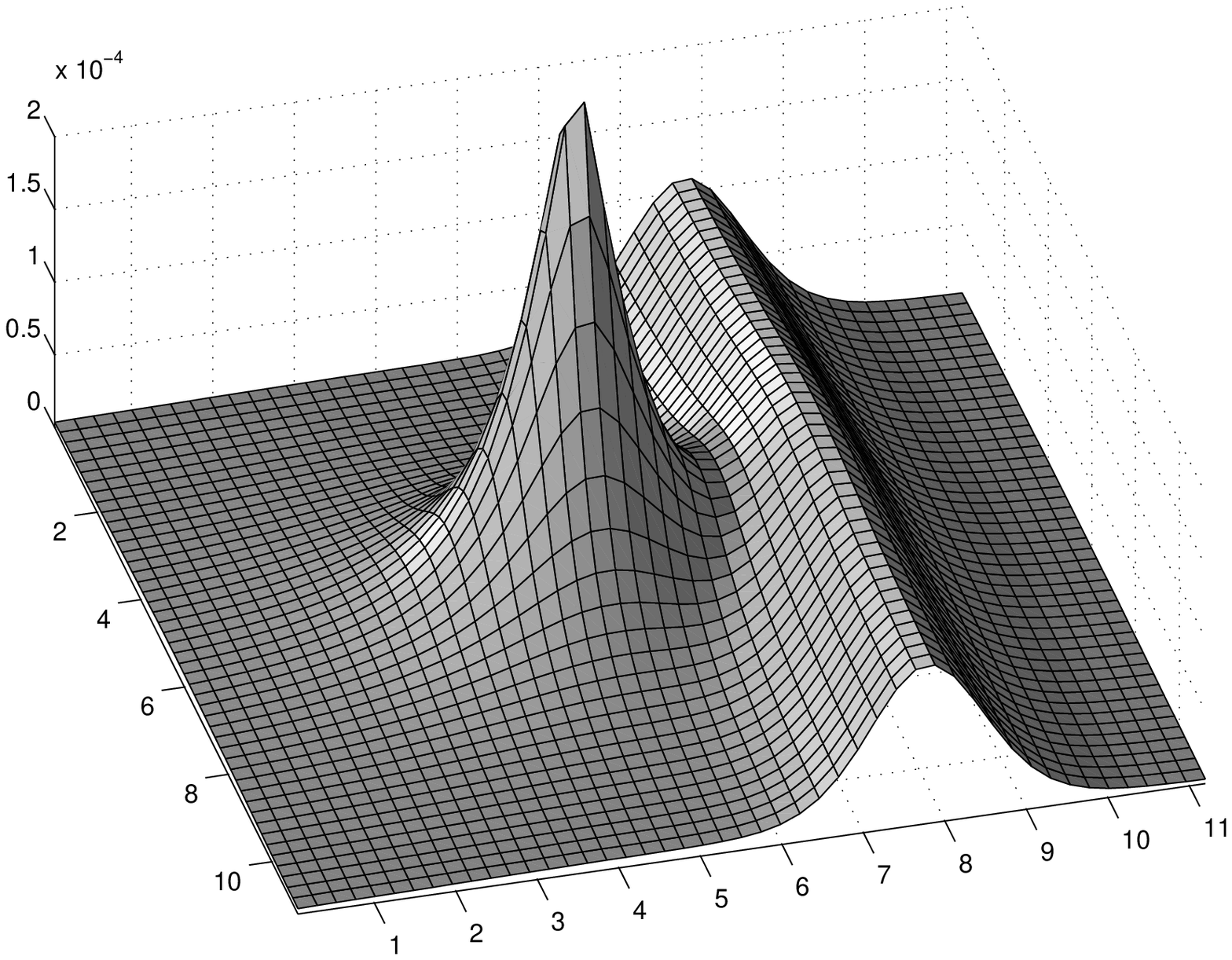} \\
\epsfysize=4.5cm 
\epsfbox{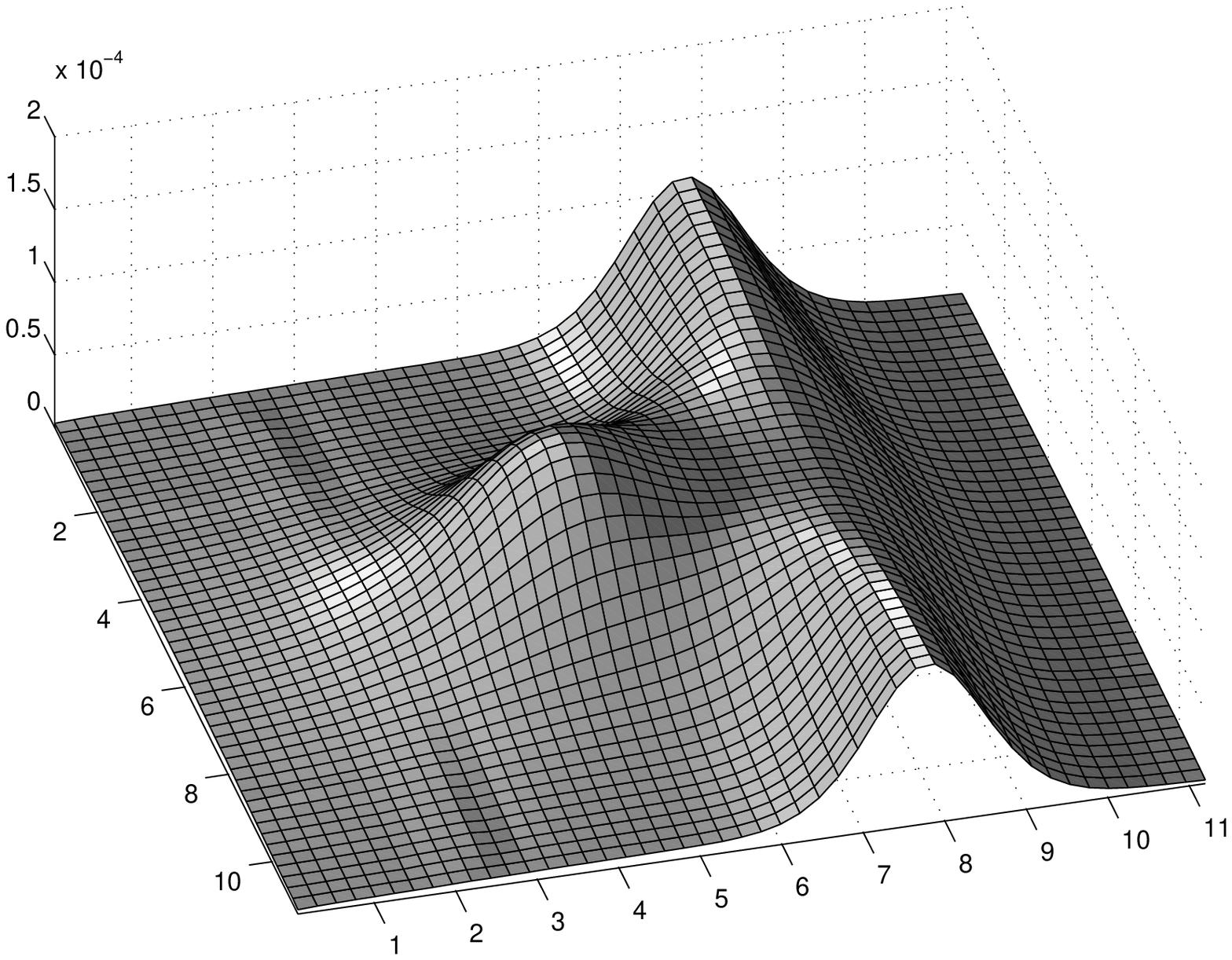} \\
\epsfysize=4.5cm 
\epsfbox{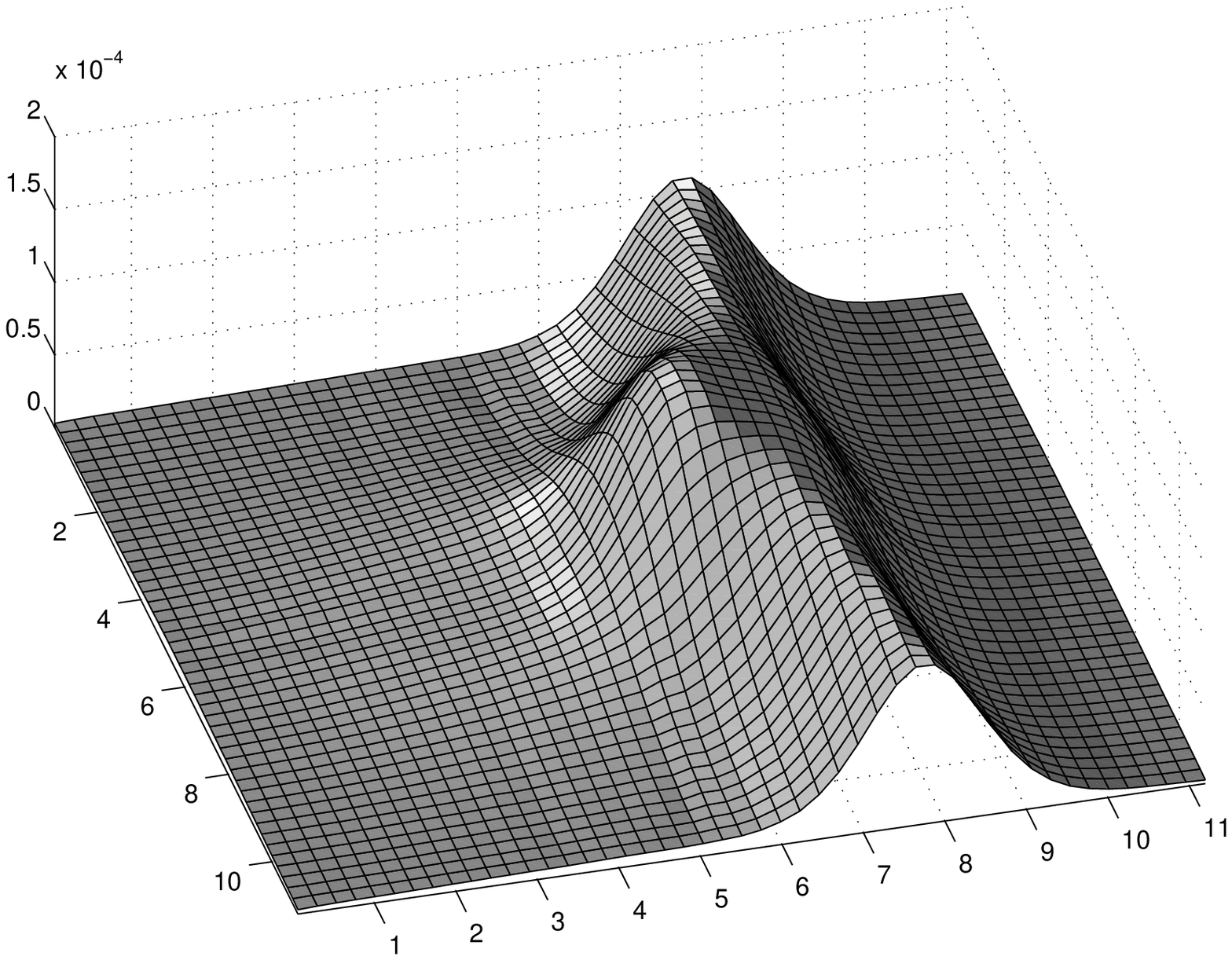} \\
\epsfysize=4.5cm 
\epsfbox{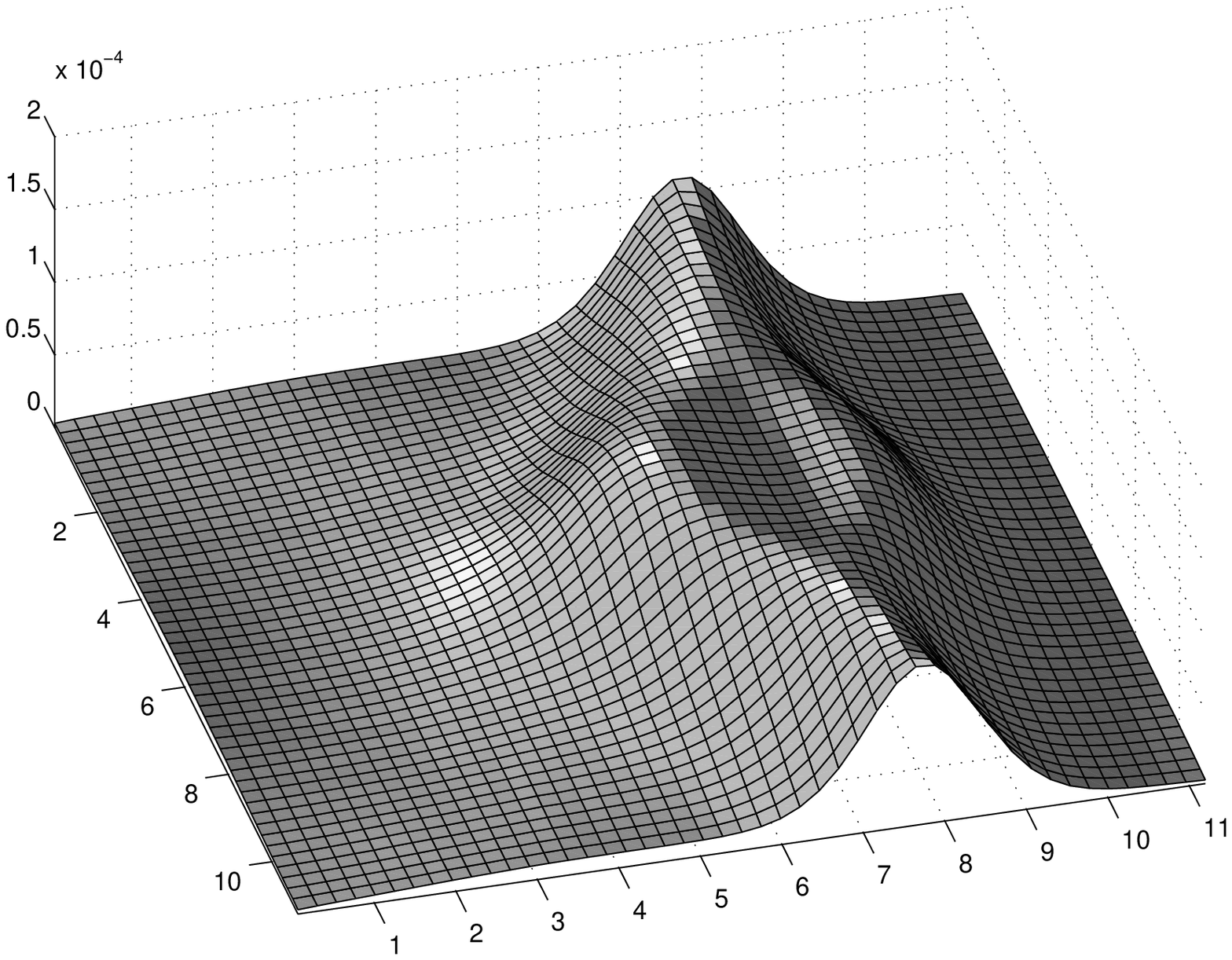} 
\end{tabular}
\end{center}
\caption{Time evolution of the energy density (vertical axis) of
  a monopole colliding with a domain wall.  The monopole is attracted
  towards the wall, and both the monopole and wall remain well defined
  after the initial interaction. The first frame shows the initial
  configuration, and subsequent frames are shown at times $1.42 n $
  ($n=1$,$2$,$3$), measured in the same (arbitrary) units in which
  $\eta_1 = \eta_2=1$.  }
\end{figure}

\begin{figure}[htbp] 
\begin{center}
\begin{tabular}{c}
\epsfysize=4.5cm 
\epsfbox{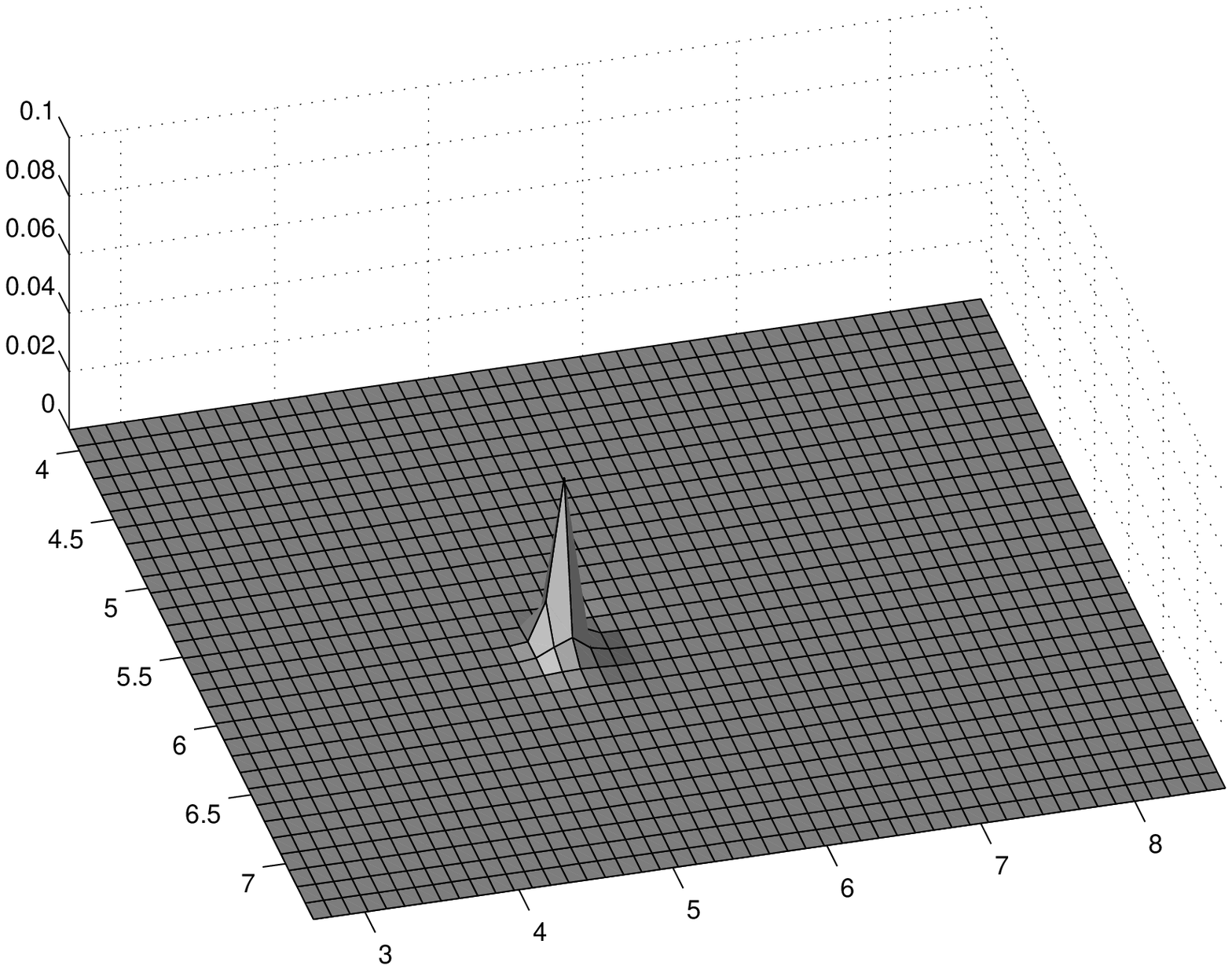} \\
\epsfysize=4.5cm 
\epsfbox{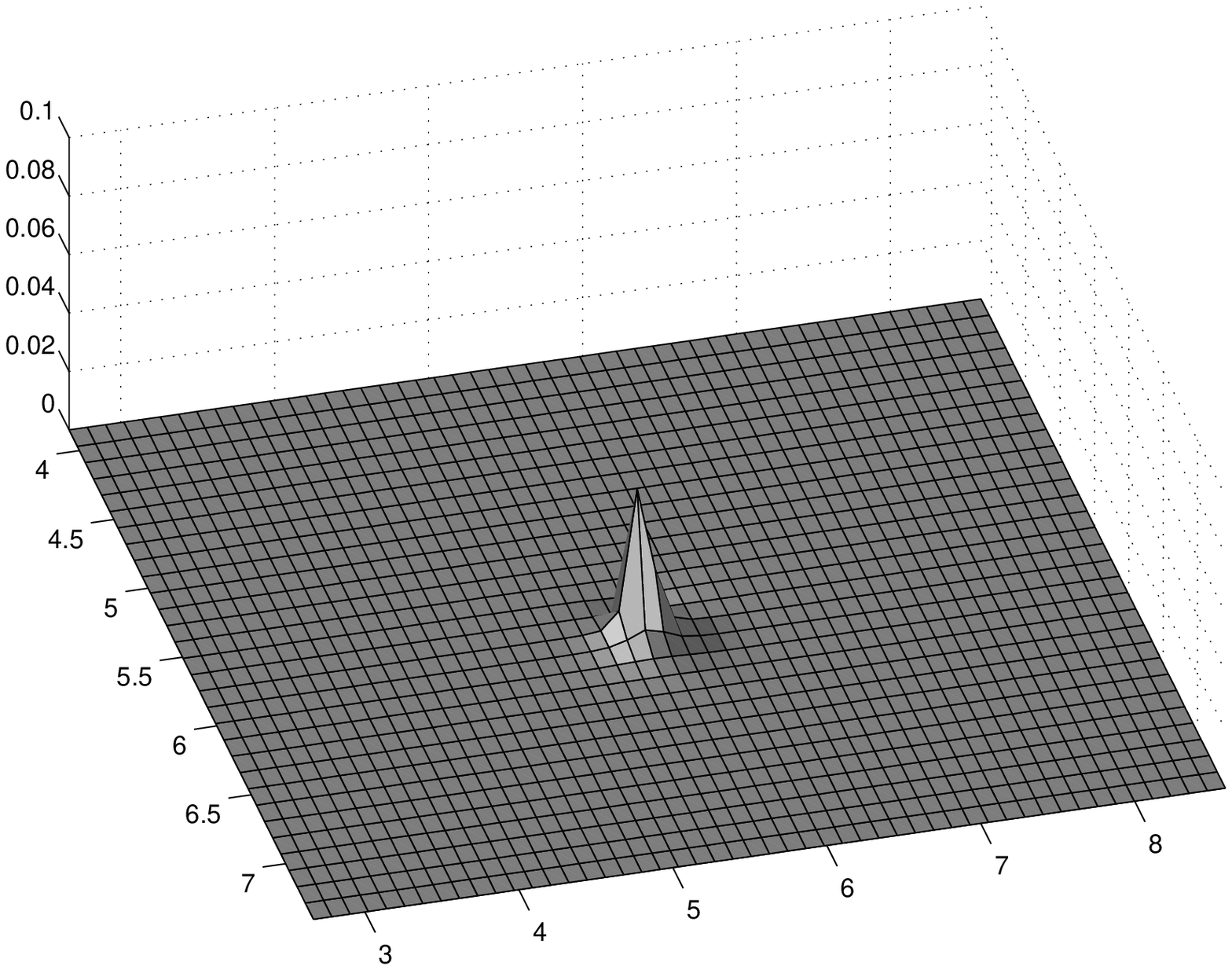} \\
\epsfysize=4.5cm 
\epsfbox{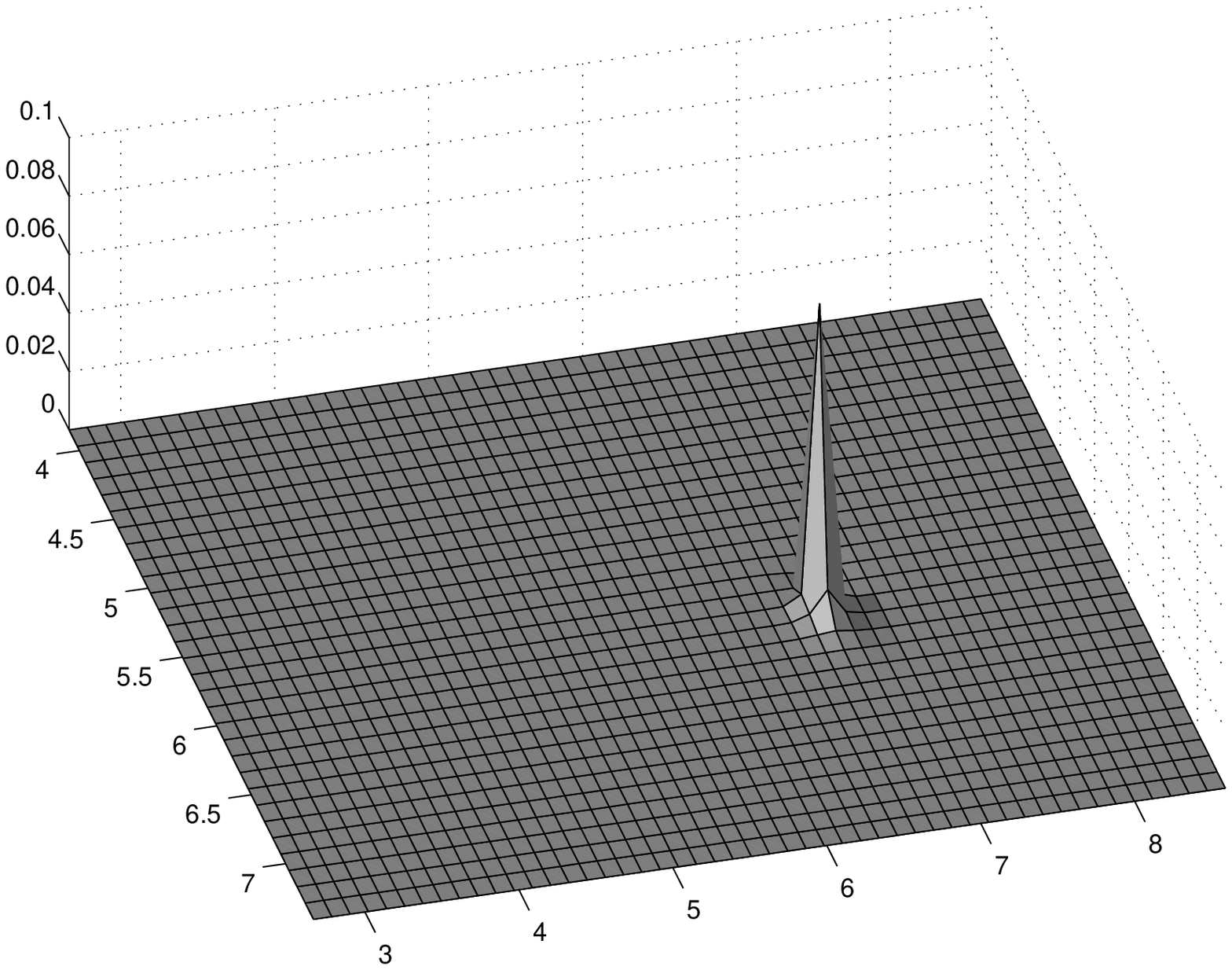} \\
\epsfysize=4.5cm 
\epsfbox{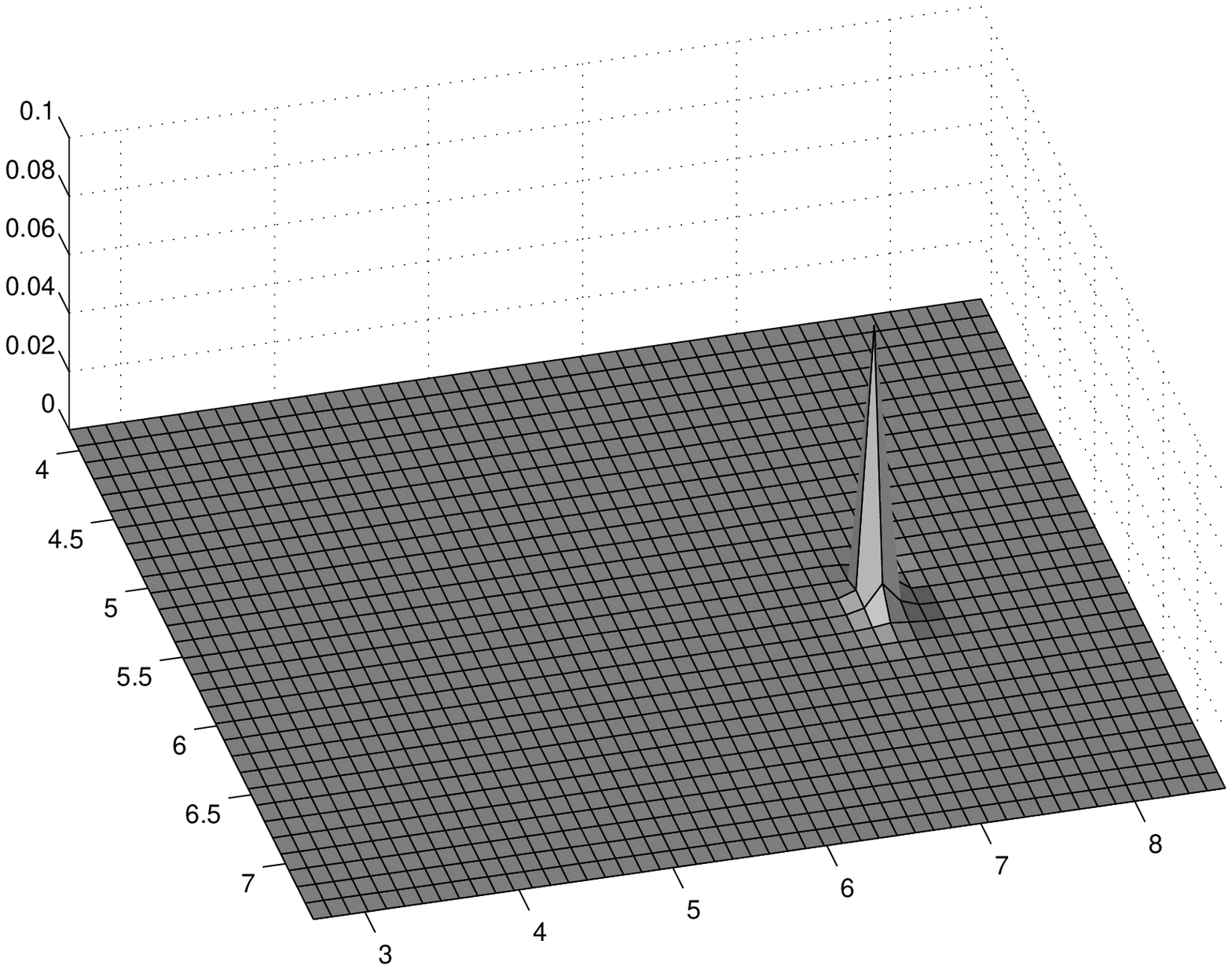} 
\end{tabular}
\end{center}
\caption{Time evolution of the winding number density of a monopole
  interacting with a domain wall. The frames are synchronous with the
  energy density plots shown in the previous figure. These plots show
  a detail from the center of the simulation volume, whereas the
  previous figure displays slices through the whole volume.  The
  height of the peak is not significant, due to the inability of the
  discrete grid upon which the solutions are calculated to accurately
  depict an approximate delta function. }
\label{figm2w} 
\end{figure}

We see from Fig. 2 that the monopole winding is initially localized at
the core of the monopole.  The charge remains localized on the wall at
the point of impact for late times and does not dissipate (unwind) on
the wall.  This differs from the evolution seen when a monopole
interacts with an embedded wall \cite{ABES}, in which case the
monopole can unwind on the wall.  Consequently, we will now examine
the differences between the two models so as to understand the lack of
unwinding with a stable domain wall, but in order to do this we must
understand the mechanism which attracts the domain wall to the
monopole.

The mechanism underlying the attraction between general monopoles and
domain walls is based on the assumption that the fields evolve to
minimize the sum of gradient and potential energies. As shown in
\cite{ABES}, the gradient energy decreases as the monopole approaches
the wall.  Another way to understand the origin of the attraction is
to focus on the change in potential energy and to associate it with a
force between the monopole and the wall.  If the force is attractive,
it behaves like a tension which induces an acceleration along the line
that connects the solitons.
 
{F}rom the form of the potential in (\ref{pot}) it is obvious that the
potential energy associated with the monopole part of the
configuration ($\phi_1 - \phi_3$) decreases with the distance between
the monopole and the domain wall, as the monopole's contribution to
the potential energy being well localized. The potential energy is
proportional to $\phi_4^2$ and thus decreases as the distance between
the monopole and the wall decreases.  However, this is a short range
force since the monopole profile function $f(r)$ approaches 1 at large
$r$. The decrease in the gradient energy as the distance between
monopole and domain wall shrinks is a longer range effect (for global
symmetries - for local symmetries the gradients are screened by the
gauge fields and hence also the force stemming from minimizing the
gradient energy is short range). This effect comes from the fact that
the volume of the region over which large gradients are localized
shrinks as the defects approach.  Note that this argument works for
general classes of defects, and adds support to the first crucial
assumption in the DLV \cite{DLV} mechanism for solving the monopole
problem.

We can understand the lack of unwinding in the present model by
studying the energetics associated with a monopole unwinding on the
domain wall. Let us assume that the monopole winding can spread out to
a distance $l$ in the plane of the domain wall. When the monopole hits
the wall its potential energy vanishes.  What is left over is the
gradient energy of both the monopole configuration (since the domain
wall is topological, there is no reason to expect the $\phi_4$
configuration to change).  As the monopole winding tries to spread
out, the $\phi_1 - \phi_3$ hedgehog configuration will be distorted
so that for most values of $x^2 + y^2 < l^2$ the $\phi_3$ component
dominates. Thus, in order to estimate the energetics associated with a
speading monopole charge, let us foucs on the contribution of the
$\phi_3$-component of the hedgehog field to the gradient
energy. Making use of the symmetry in the plane of the wall, the
energy $E$ can thus be approximated by integrating the gradient energy
density of the $\phi_3$-component of the monopole integrated over all
$z$ out to a distance $l$ in the plane of the wall.  %
\be
E > \pi l^{2}\int_{w}^{\infty} (\frac{\partial \phi}{\partial z})^{2}
dz \sim \pi l^{2} \int_{w}^{\infty} \frac{1}{z^{2}\pi^{2}} dz  
\ee
where $w$ is the thickness of the wall along the $z$ axis. Thus,
\be
E> \frac{1}{\pi}l^{2}\eta^{2}w^{-1} \,.
\ee

Hence we see that the energy grows as $l^{2}$. We can obtain an
effective force by taking a derivative with respect to $l$.  What we
see is a restoring force $F$ which acts to keep the monopole winding
charge confined to a localized region of the domain wall.
\be 
F = -kl \,\,\,\, \rm{with} \,\,\,\, k = \frac {1}{\pi} \eta^{2} w^{-1} \, .
\ee 
 
We conclude that this ``restoring force'' is the basis of the
non-unwinding of the monopole on the domain wall.

\section{Conclusions}

We have studied the interaction between monopoles and domain walls in a
scalar field theory with symmetry group $G = O(3) \times Z_2$ symmetry
group spontaneously broken to $H = O(2)$, a theory which admits both
stable monopoles and domain walls. We demonstrated the existence of
one of the two mechanisms which was crucial for the proposed solution
of the monopole problem of \cite{DLV}, namely the attractive form
between monopoles and domain walls.  However, in contrast to the case
of global monopoles interacting with embedded walls \cite{ABES} we
found that the charge of the global monopole will not unwind on the
surface of the domain wall.  The gradient energy associated with
delocalizing the monopole topological charge yields a tension which
keeps the winding confined to a local region on the wall.  Hence, we
conclude that global monopoles can not unwind unless there are gauge
fields to cancel the tension associated with the gradients of the
scalar fields. Therefore, the lack of unwinding of global defects on
domain walls is a universal behavior.

\section*{Acknowledgements} We thank Bill Unruh for a very useful
discussion.  Computational work in support of this research was
performed at the Theoretical Physics Computing Facility at Brown
University.  This work has been supported in part by the US Department
of Energy under Contract DE-FG02-91ER40688, TASK A, and one of us (SA)
acknowledges support from the US Department of Education under the
GAANN program.

\end{document}